# Experimental Characterization of the Pyridine:Acetylene Co-crystal and Implications for Titan's Surface

Ellen C. Czaplinski,* Tuan H. Vu, Morgan L. Cable, Mathieu Choukroun, Michael J. Malaska, and Robert Hodyss

**ABSTRACT:** Titan, Saturn's largest moon, has a plethora of organic compounds in the atmosphere and on the surface that interact with each other. Cryominerals such as co-crystals may influence the geologic processes and chemical composition of Titan's surface, which in turn informs our understanding of how Titan may have evolved, how the surface is continuing to change, and the extent of Titan's habitability. Previous works have shown that a pyridine:acetylene (1:1) co-crystal forms under specific temperatures and experimental conditions; however, this has not yet been demonstrated under Titan-relevant conditions. Our work here demonstrates that the pyridine:acetylene co-crystal is stable from 90 K, Titan's average surface temperature, up to 180 K under an atmosphere of $N_2$. In particular, the co-crystal forms via liquid–solid interactions within minutes upon mixing of the constituents at 150 K, as evidenced by distinct, new Raman bands and band shifts. X-ray diffraction (XRD) results indicate moderate anisotropic thermal expansion (about 0.5–1.1%) along the three principal axes between 90–150 K. Additionally, the co-crystal is detectable after being exposed to liquid ethane, implying stability in a residual ethane "wetting" scenario on Titan. These results suggest that the pyridine:acetylene co-crystal could form in specific geologic contexts on Titan that allow for warm environments in which liquid pyridine could persist, and as such, this cryomineral may preserve the evidence of impact, cryovolcanism, or subsurface transport in surface materials.

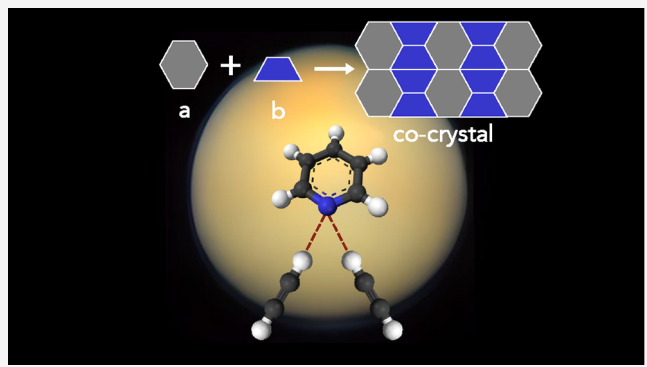

## 1. INTRODUCTION

Titan, Saturn's largest moon, contains a multitude of organic molecules in the atmosphere and on the surface. Solar radiation and energetic protons from Saturn's magnetosphere provide a unique environment, generating a photochemical cascade where $N_2$ and $CH_4$ dissociate, ionize, and recombine to create simple (acetylene, ethane, hydrogen cyanide, and other small nitriles and hydrocarbons) and complex (>10,000 Da) organic molecules as they travel through Titan's atmosphere.[1−5] These organic compounds are delivered to the surface where they likely comprise the majority of surface materials and are subjected to transport by eolian, fluvial, and even lacustrine processes by the primarily liquid methane phase of Titan's hydrologic cycle. Here, we studied two Titan-relevant compounds, acetylene and pyridine, to determine whether they form a co-crystal (a type of molecular mineral) when allowed to interact under Titan atmospheric and surface conditions. Co-crystals can exhibit unique chemical and physical properties compared to their pure molecular constituents and as such, can be good indicators of geologic or geochemical processes occurring on Titan's surface.

Acetylene ($C_2H_2$) is one of the primary photochemical products in Titan's atmosphere[6] ($2.8 \times 10^{-4}$ mole fraction at 1100 km;[7] Vuitton et al.) that likely forms through a multistep process of photolysis of methane and ethylene (Table 1[4,8]); it has been tentatively identified in the atmosphere and on the surface from spectral analysis (e.g.,[9−11]). As a solid, acetylene has two crystalline phases: a low-temperature orthorhombic phase (below 133 K) and a high-temperature cubic phase (133–193 K).[12] Because of Titan's low surface temperature (89–94 K), the orthorhombic phase of acetylene is the expected form on the surface.

Pyridine ($C_5H_5N$) is a simple nitrogen heterocycle, a class of molecules that have been identified in meteoritic organic matter,[13−17] and nitrogen-based heterocycles are fundamental to Earth-based life.[18,19] Additionally, the enhanced stability of aromatics, including pyridine, makes them good candidates for detection by both in situ and sample return missions.[19] Although pyridine has not been directly detected in Titan's atmosphere, when in the presence of HCN (which has been detected and routinely observed in Titan's atmosphere[20−24]), acetylene polymerization may produce N-heterocycles including pyridine.[18,25] Further, a ring expansion reaction (gas phase) between electron-deficient methyl carbyne (CH) and pyrrole ($C_4H_5N$) (an N-heterocycle) directly produces pyridine.[26] Upper limits on pyrrole in Titan's atmosphere have been inferred at <4.0 × 10$^{-8}$ in the stratosphere using Voyager data[27] and <3.0 × 10$^{-7}$ in the thermosphere using the Cassini Ion and Neutral Mass Spectrometer (INMS).[28] The existence of a few tenths of a ppm of pyridine in the upper atmosphere (4.0 × 10$^{-7}$ mole fraction at 1100 km) has been inferred from ion densities at m/z = 80 and 94 from a previous photochemical model.[7] Additionally, this photochemical model suggests two unidentified N-containing species, of which pyridine is a probable candidate.[7] A 2$\sigma$ upper limit of ∼1.15 ppb has been reported for pyridine in Titan's upper atmosphere (constant profile above 300 km).[29]

Table 1. Formation Reactions of Acetylene and Pyridine in Titan's Atmosphere, Density and Altitude in Titan's Atmosphere, and the Mole Fraction

| species | formula | formation reaction (s) | density (g/cm$^{-3}$) | mole fraction in Titan's atmosphere |
|---|---|---|---|---|
| acetylene | $C_2H_2$ | $C_2H_4 + h\nu \rightarrow C_2H_2 + 2H/H_2$ | 0.61[a] | 3.1 × 10$^{-4}$[b] |
| pyridine | $C_5H_5N$ | $CH + C_4H_5N \rightarrow C_5H_5N + H$[d] | 1.149[c] | 3.0 × 10$^{-7}$[d] |

[a]From 37, 194 K. [b]From 28, the closed source neutral (CSN) mode of the Cassini Ion and Neutral Mass Spectrometer (INMS) at 1077 km atmospheric height. [c]From 19, 130 K, crystalline. [d]From 38, the inferred mole fraction at 1100 km atmospheric height.

Co-crystals are compounds with a set stoichiometric ratio; they are stable structures held together by relatively weak intermolecular interactions (e.g., London dispersion forces and pi bonding).[30] These weak intermolecular interactions have proven important in cryogenic environments such as the surface of Titan, leading to molecular minerals that may be stable for even geologic timescales. Previously, several Titan-relevant co-crystals have been identified experimentally from observing spectral shifts in both Raman and Fourier-transform infrared (FTIR) spectra, concurrent with changes in X-ray diffraction (XRD) patterns and sample morphology. Since 2014, seven organic co-crystals have been reported and characterized under Titan-relevant experimental conditions, including another nitrile:acetylene co-crystal (acetonitrile:acetylene (1:2)[31]). Many of these previous co-crystal studies included acetylene, a highly reactive molecule owing to its carbon−carbon triple bond and high energy of formation.[32] Currently, there is no single predictor as to if a molecular system will successfully form a co-crystal; however, when acetylene is one of the components, the system is less favorable if the non-acetylene molecule has a low-energy structure.[33] Interestingly, when acetylene and pyridine interact under specific temperatures and molar ratios, a co-crystal can form.[33] For example, Kirchner et al. condensed acetylene at 77 K in a 0.3 mm diameter quartz capillary filled with pyridine and pressurized up to ∼100 bar while utilizing an optical heating and crystallization device to grow the crystal.[33] However, it is important to note that pyridine would have a relatively low abundance in Titan's atmosphere (if present). It is uncertain whether pyridine and acetylene would have the opportunity to interact as two pure compounds, given the likelihood that surface materials on Titan are complex mixtures comprised of additional organic compounds.

Here, we report that the pyridine:acetylene (1:1) system forms a stable co-crystal under Titan-relevant temperatures (90 to 180 K). We note that this temperature range correlates with Titan's subsurface and atmospheric temperatures, as the tentative subsurface ocean may reach temperatures above 250 K[34,35] and the atmosphere reaches temperatures >150 K above 100 km altitude.[36] These results add to the body of knowledge on this rapidly expanding field of Titan cryomineralogy, which can help discern the surface-scale composition and inform large-scale geologic processes on Titan.

## 2. EXPERIMENTAL TECHNIQUES

**2.1. Sample Preparation.** Acetylene (Airgas, Inc., industrial grade, dissolved in acetone) was passed through a purifier (Micro Torr MC400−404F, SAES Pure Gas, Inc.) to remove particles <0.003 μm and organic impurities to <1 pptV (ppt by volume) prior to use, as verified by the absence of Raman spectral features at 787, 1710, and 2922 cm$^{-1}$ of acetone. After purification, acetylene was injected into a gas sample bag (0.7 L 2 mil Tedlar film, single polypropylene septum fitting, SKC, Inc.) for subsequent deposition. For Raman experiments, a 50 μL aliquot of pyridine (Sigma-Aldrich, ≥99.0%) was deposited onto one of two depressions (or wells) of a 5 mm thick, 2-well microscope slide at 273 K within a liquid nitrogen-cooled optical cryostage (LTS 350, Linkham Scientific Instruments, Ltd.). The pyridine aliquot was deposited on the well opposite to the liquid nitrogen-cooled area of the stage to allow for it to condense from the headspace vapor to the lower temperature well of the slide as the temperature decreased. Acetylene was subsequently condensed for ∼5 to 10 s from the gas phase via the sample bag into the cryostage at each temperature increment, starting at ∼250 K. This technique allowed for a ratio of pyridine:acetylene that was optimal for co-crystal formation. The cryostage was cooled in increments of 10 K every 2 min under an atmosphere of $N_2$ until Titan's surface temperature (∼90 K) was reached. We note that these experiments were performed under a $N_2$ atmosphere of 1 bar, whereas Titan surface pressure is 1.5 bar. A schematic of the experimental setup for Raman spectroscopic measurements is depicted in Figure S1.

For powder X-ray diffraction (XRD) experiments, an ∼8 μL aliquot of pyridine was deposited into an open-ended borosilicate capillary (0.7 mm internal diameter). The capillary was then mounted and aligned on the goniometer sample attachment of the XRD. The open end of the capillary was attached to a custom-built system for introducing gases (in this case, acetylene) into the capillary, which allows for precise manipulation and deposition of the analyte gas.[39] The system is comprised of two valves and a flowmeter which are connected to an 8 cm long polyamide-coated silica capillary tube (360 μm outside diameter, 100 μm inside diameter) through a standard 1/8″ Swagelok elbow, which is mounted to a manual XYZ micromanipulator.[39] The silica capillary was slowly directed inside of the borosilicate capillary to prepare for acetylene deposition. Following the nitrogen purge, the acetylene gas flow into liquid pyridine was initiated at room temperature; the sample temperature was gradually lowered in ∼10 K increments using a liquid nitrogen-cooled Oxford Cryosystems Cryostream 800 (temperature control to within ±1 K) until the mixed sample solidified at ∼186 K. For ethane wetting (mixing) experiments, the sample was cooled to 110 K after the co-crystal was verified to form within the capillary at 180 K. A 1 L Tedlar gas sample bag was filled with gaseous ethane, which was subsequently condensed through the custom-built gas introduction system and into the capillary so the liquid ethane could mix with the pyridine:acetylene co-crystal sample. A schematic of the experimental setup for the XRD measurements is depicted in Figure S2. No unexpected or unusually high safety hazards were encountered in either the micro-Raman or the XRD experiments.

**2.2. Raman Spectroscopy.** Raman spectroscopy is an important method for studying a variety of materials including co-crystals, as it provides information about both the composition and the chemical environment of the molecules being studied. The co-crystal formation is typically identified by frequency shifts, splitting and merging of vibrational modes, or sharpening of peaks compared to spectra of the pure components. Raman measurements were performed using a high-resolution confocal dispersive micro-Raman spectrometer (Horiba Jobin-Yvon LabRam HR).

After both compounds were deposited, they were observed with the micro-Raman spectrometer through the optical window of the cryostage, which was mounted onto an XYZ motorized translation stage (Märzhäuser Wetzlar) underneath the Olympus BXFM objective turret of the micro-Raman spectrometer. The sample was observed continuously under various levels of magnification (4×, 10×, 50×) during the experiment. Raman spectra were collected at 0.4 cm$^{-1}$ per pixel resolution using an 1800 grooves/mm grating or 1.7 cm$^{-1}$ resolution using a 600 grooves/mm grating. All samples were excited by a neodymium-doped yttrium aluminum garnet (Nd:YAG) laser that was frequency-doubled to 532 nm, with an output power of 50 mW. The silicon 520.7 cm$^{-1}$ peak was used for frequency calibration. Spectra were collected with acquisition times of 45−90 s, depending on the signal strength of the particular sample. Thermal stability studies were performed by warming the sample in 10 K increments and obtaining Raman spectra after a 2 min equilibration time at each temperature point.

**2.3. Powder X-ray Diffraction.** Powder XRD is a useful tool for characterizing the co-crystal structure, phase, and thermal expansion/contraction. XRD measurements were performed using a Bruker D8 Discover Da Vinci X-ray diffractometer. The co-crystal formation was confirmed immediately after sample solidification via the identification of characteristic peaks in the XRD pattern. The silica capillary was withdrawn and the borosilicate capillary was rapidly flame-sealed to isolate the sample from the atmosphere during XRD measurement. Powder XRD patterns were then collected from 90 to 150 K at intervals of 10 K with 10 min of equilibration at each temperature point (2 s per step with a 2θ angular resolution of 0.02°, which resulted in ∼2 h for each pattern) using a Cu Kα X-ray source (λ = 1.5406 Å) and a linear energy-dispersive LynxEye XE-T one-dimensional (1D) detector. Additional ethane mixing (wetting) experiments were performed with ethane following co-crystal confirmation. All data were analyzed using Bruker's Diffrac TOPAS suite (version 6).

## 3. CO-CRYSTAL FORMATION

We compared the co-crystal spectra with pure acetylene, pure pyridine, and the acetylene clathrate hydrate, which has similar bands in the C≡C stretching region (Figures 1−4 and Tables

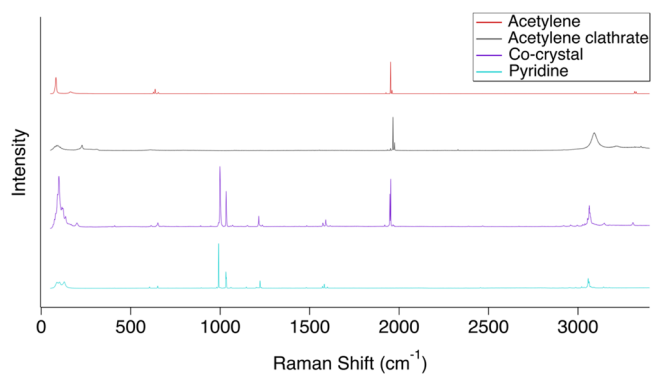

**Figure 1.** High-resolution Raman spectra of solid acetylene (red), the acetylene clathrate (gray, 4× scale), the pyridine:acetylene (1:1) co-crystal (purple), and solid pyridine (blue). All spectra were collected at 90 K. The acetylene clathrate spectrum is by Vu et al. (2019).[41] Spectra are vertically offset for clarity.

**Table 2. Experimental Raman Shifts of Acetylene after Co-crystal Formation (90 K), Compared to Reported Raman Band Centers for Pure Acetylene and the Acetylene Clathrate (90 K)**

| | Raman shift (cm$^{-1}$) | | | | |
|---|---|---|---|---|---|
| | pure acetylene | | acetylene clathrate | co-crystal | Δν between pure acetylene and co-crystal[d] |
| vibrational mode[a] | reported[b] | this work | reported[c] | this work | this work |
| $\nu_4$ (C≡C–H bend) | 628.5 | 626.7 | | | |
| | 638.5 | 636.5 | | | |
| | 659.5 | 654.5 | | | |
| $\nu_2$ (C≡C stretch) | 1951.5 | 1951.8 | | 1948.3 | −3.5 |
| | | | | 1953.1 | 1.3 |
| | 1960.5 | 1960.3 | 1965.5 | 1966.0 | 5.7 |
| | | | 1974.4 | 1972.5 | |
| water ice (bonded O–H stretch) | | | 3089.3 | | |
| $\nu_1$ (C–H stretch) | 3314.5 | 3317.0 | | 3307.1 | −9.9 |
| | 3323.5 | 3325.1 | | | |

[a]Lattice vibrational modes are listed in Table S1. [b]From 42. [c]From 41. [d]Positive value of Δν indicates a blue shift; a negative value indicates a red shift.

2 and 3). After acetylene was deposited with the pyridine sample (∼190 K) and cooled to ∼150 K, redshifts (bathochromic) and blueshifts (hypsochromic) up to ∼15 cm$^{-1}$ are observed in the most prominent vibrational modes of each pure molecule (Tables 2 and 3) and are described in the following sections. New bands are present in the co-crystal

**Table 3. Experimental Raman Shifts of Pyridine after Co-crystal Formation (90 K), Compared to Reported Raman Band Centers for Pure Pyridine (90 K)**

| | Raman shift (cm$^{-1}$) | | | |
|---|---|---|---|---|
| | pure pyridine | | co-crystal | $\Delta\nu$ between pure and co-crystal |
| vibrational mode[a] | reported[b] | this work | this work | this work[c] |
| $\nu_{6a}$ (in-plane ring bend)[43] | 603 | 605.5 | 613.8 | 8.3 |
| $\nu_{6b}$ (in-plane ring bend)[43,44] | 651 | 650.8 | 651.8 | 1 |
| $\nu_{21}$[44] | 893 | 894.4 | | |
| | | 980.7 | | |
| $\nu_1$ (C−C ring stretch)[43] | 992 | 991.2 | 998.6 | 7.4 |
| $\nu_8$ (in-plane ring bend)[44] | 1032 | 1032 | 1033.4 | 1.4 |
| | | 1034.8 | | |
| | | 1060.1 | | |
| $\nu_{15}$ (in-plane H)[43] | 1145 | 1146.5 | 1153.3 | 6.8 |
| | 1204 | 1201.1 | | |
| $\nu_{16}$ (in-plane H bend)[44] | 1222 | 1222.7 | 1215.9 | −6.8 |
| | | 1230.7 | 1236 | 5.3 |
| $\nu_{8b}$ (C−C ring stretch)[43] | 1572 | 1571.7 | 1573.8 | 2.1 |
| $\nu_{8a}$ (C−C ring stretch)[44] | 1582 | 1581.5 | 1589.6 | 8.1 |
| | | 1599.8 | 1615.1 | 15.3 |
| $\nu_{12}$[44] | 3021 | 3020.2 | 3026 | 5.8 |
| | | | 3038.6 | |
| $\nu_2$ (C−H stretch)[43,44] | 3054 | 3056.3 | 3053.9 | −2.4 |
| | | 3062 | 3063.4 | 1.4 |
| | | 3069.5 | 3066.8 | −2.7 |
| | | 3142.9 | 3146.5 | 3.6 |
| | | 3156.5 | | |
| | | 3174.7 | | |

[a]Lattice vibrational modes are listed in Table S1. [b]From 43. [c]Positive value of $\Delta\nu$ indicates a blue shift; a negative value indicates a red shift.

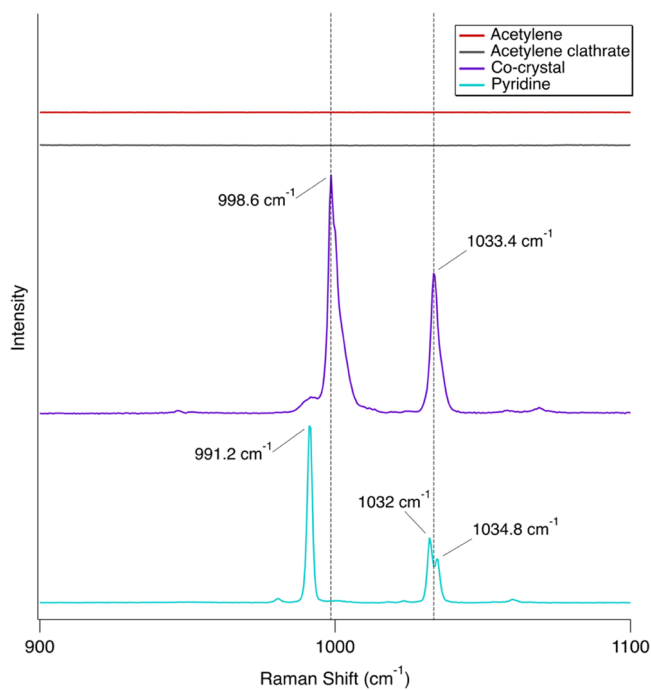

**Figure 2.** Inset of high-resolution Raman spectra from Figure 1 showing the $\nu_1$ (991.2 cm$^{-1}$; C−C ring stretch) and $\nu_{12}$ (1032 and 1034.8 cm$^{-1}$; in-plane ring bend) bands of pyridine compared to the pyridine:acetylene (1:1) co-crystal. From top to bottom: solid acetylene (red), the acetylene clathrate (gray, 4× scale), the pyridine:acetylene (1:1) co-crystal (purple), and solid pyridine (blue). Spectra are scaled for clarity by the same multipliers as in Figure 1. All spectra were collected at 90 K. The blueshifts of the pyridine $\nu_1$ band (dashed vertical lines) are unique to the co-crystal. Notice the merging of the co-crystal band at 1033.4 cm$^{-1}$ when compared to the associated split pyridine $\nu_{12}$ bands. Pure acetylene and the acetylene clathrate have no features in this region but are included for completeness. Spectra are vertically offset for clarity.

spectrum—a common aspect of co-crystal formation. These new bands are associated with a change in the molecular environment when the co-crystal forms, as compared to the molecular environment of the two pure species. Additionally, Figure S4 shows a spectrum of the pyridine trihydrate compared to pure pyridine and the pyridine:acetylene (1:1) co-crystal, confirming the distinction of the co-crystal spectrum.

The lattice vibrations arise from the translational and rotational motion of the molecules in the solid. New features are observed in the low-frequency lattice vibration modes (~50−200 cm$^{-1}$) (Figure S3 and Table S1) at 115.4, 121.7, and 199.4 cm$^{-1}$. Band splitting and shifting are also observed. The overall band shape broadened and increased in intensity upon co-crystal formation.

The C−C ring stretching occurs in pyridine when the bonds that connect the C atoms in the molecule lengthen. The in-plane bending occurs when C−H bonds bend in the plane of the pyridine aromatic ring. Upon co-crystal formation, blue shifts occurred in the $\nu_1$ and $\nu_{12}$ pyridine bands (Figure 2). Broadening of both bands and merging of the split $\nu_{12}$ pyridine band (1033.4 cm$^{-1}$) also occurred after co-crystal formation (Figure 2).

The C≡C stretching in acetylene occurs when the C−C distances change as the bond stretches and compresses. New bands observed in the co-crystal spectrum at 1948.3, 1953.1, and 1966 cm$^{-1}$ are a clear indicator of co-crystal formation (Figure 3), similar to those seen by Cable et al. (2020)[31] for the acetonitrile:acetylene co-crystal. Specifically, the new band at 1953.1 cm$^{-1}$ is associated with how the pyridine and acetylene molecules are arranged within the co-crystal environment (refer to Section 4). Note that the band at 1974.4 cm$^{-1}$ in the acetylene clathrate spectrum (Figure 3) is from acetylene in the gas phase as sublimated acetylene filled the headspace (similar to what occurred with the butane:acetylene co-crystal[40]).

The C−H stretching region shown is comprised of C−H vibrational motions for both acetylene and pyridine. Acetylene shows two sharp peaks at 3317 and 3325.1 cm$^{-1}$, while the co-crystal has a single, broader peak at 3307.1 cm$^{-1}$ (Figure 4); the emergence of this single, broad peak indicates co-crystal formation, as reported by Cable et al. (2020).[31] Changes to the crystal structure of the sample is evidenced by the increased broadening and intensity of pyridine bands near the peak at 3063.4 cm$^{-1}$. This region of the spectrum is complex, with many overlapping features, so no comprehensive analysis of the changes was attempted.

**3.1. Sample Morphology.** As liquid pyridine accumulates in the empty slide well, it initially forms droplets (~15 to 100 $\mu$m in diameter) (Figure 5, top left). The droplets increase in size as the temperature decreases. Acetylene crystallizes inside the pyridine matrix (top right panel in Figure 5; the dark-toned texture indicates acetylene crystallization within the light-toned pyridine matrix) when acetylene is allowed to condense within the cryostage at 185 K (Figure 5, middle). We note that Figure 5 was taken during acetylene deposition to depict an example of acetylene crystallization within pyridine and also to visually compare this crystallization to the pure pyridine droplets (top right panel in Figure 5, bottom right of the image). When the sample is cooled to Titan temperatures after acetylene condensation, certain regions of the sample become dark (lower albedo) and form an irregular texture (Figure 5, right), surrounded by lighter areas of pure pyridine.

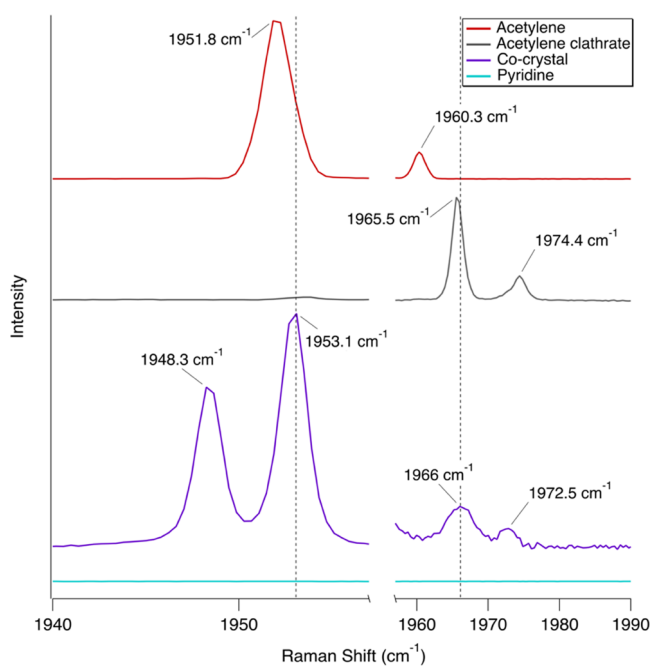

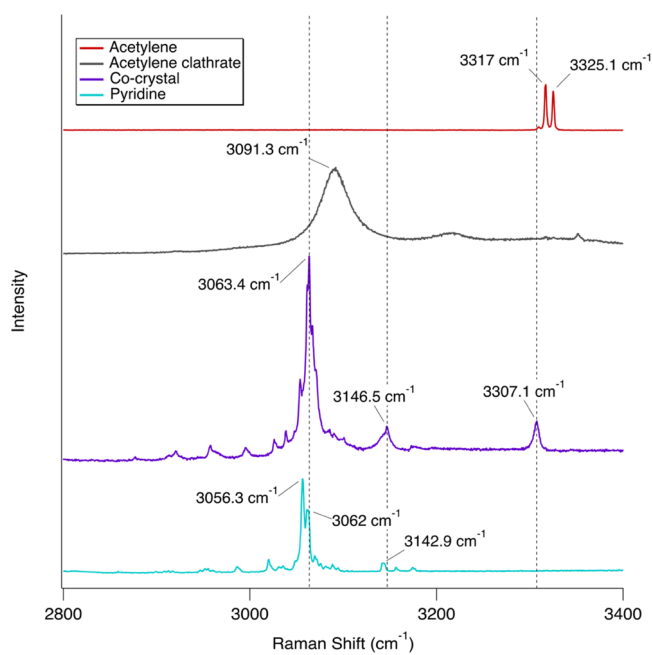

**Figure 3.** Inset of high-resolution Raman spectra from Figure 1 showing the $\nu_2$ (1951.8 and 1960.3 cm$^{-1}$; C≡C stretch) bands of acetylene compared to the pyridine:acetylene (1:1) co-crystal. The *x*-axis was split so the spectra on the right side of the break could be scaled for clarity. The scale left of the break: acetylene (4×), acetylene clathrate (4×), and co-crystal (4×). The scale right of the break: acetylene (6×), acetylene clathrate (10×), and co-crystal (20×). All spectra were collected at 90 K. A new band in the co-crystal spectrum at 1948.3 cm$^{-1}$ and the blueshift of the 1960.3 cm$^{-1}$ band to 1966 cm$^{-1}$ (dashed vertical lines) are clear indicators of co-crystal formation. Note that the acetylene clathrate band at 1974.4 cm$^{-1}$ is from acetylene in the gas phase as sublimated acetylene filled the headspace. Pure pyridine has no features in this region but is included for completeness. Spectra are vertically offset for clarity.

**Figure 4.** Inset of high-resolution Raman spectra from Figure 1 showing bands in the C−H stretching region compared to the pyridine:acetylene (1:1) co-crystal spectrum. Spectra are scaled for clarity as follows: acetylene (2×) and acetylene clathrate (2×). All spectra were collected at 90 K. The lack of splitting in the co-crystal band at 3307.1 cm$^{-1}$ when compared to the associated acetylene bands (3317 and 3325.1 cm$^{-1}$) indicates co-crystal formation. Formation of the co-crystal is also evidenced by changes in shape and intensity of bands near the peak at 3063.4 cm$^{-1}$. Spectra are vertically offset for clarity.

## 4. THERMAL STABILITY AND EXPANSION

The pyridine:acetylene co-crystal forms within minutes at ∼150 K. We observe supercooling in these experiments, which caused pyridine to persist as a glass (liquid-like state) below its typical freezing point of 231.6 K. Raman spectra indicate that the co-crystal is stable from Titan surface temperatures (∼90 K) to 180 K (Figure 6) and dissociates at 190 K, which is consistent with the sublimation point of acetylene (∼−84 °C/189 K).[45] This stability range is distinct from the acetylene clathrate, which is stable up to 233 K.[41] Pyridine features persist above 190 K, indicating that pyridine reverts to its pure crystalline form once the co-crystal dissociates.

The pyridine:acetylene co-crystal adopts a monoclinic structure consisting of one pyridine molecule opposed to two half-molecules of acetylene via a chain of hydrogen bridges (C−H⋯N). This orientation gives the co-crystal a 1:1 composition.[33] The XRD pattern of the pyridine:acetylene co-crystal was studied as a function of temperature between 90−150 K, with a pattern at 110 K shown in Figure 7 as an example. Distinctive diffraction peaks due to the co-crystal formation (e.g., at 10.99, 19.41, 20.13, 20.25°) were immediately apparent once the pyridine−acetylene mixture was cooled. The pattern at each temperature step was analyzed via the Pawley method using the space group $P2_1/n$ for the co-crystal (in accordance with previous results[33]), $Pna2_1$ for the unreacted pyridine,[46] and $Pbca$ to account for some amount of the pyridine trihydrate[46] that was also formed over the long-duration experiment. Table 4 lists the refined lattice constants and unit cell volume of the pyridine:acetylene co-crystal from 90−150 K.

To illustrate the variation of these values with temperature, we have used the web-based program PASCal[47] to calculate the percent change in length along the principal axes and unit cell volume, as shown in Figure 8. The co-crystal is observed to exhibit a positive thermal expansion with moderate anisotropy, up to 1.1% along the X3 axis, 0.8% in X2, and ∼0.6% in the X1 direction. This behavior is most likely due to relatively strong N⋯H−C interactions throughout the monoclinic structure (graphically presented in Figure 9), where each pyridine atom is stabilized by two acetylene molecules at distances of 2.464 and 2.528 Å. The former, slightly shorter N⋯H contact, can be found to reside mostly along the *a* and *c* axes (which lie in the plane formed by X1 and X2), thereby leading to the smaller thermal expansion in these directions relative to X3 (which coincides with the *b* axis). A similar anisotropic thermal expansion behavior has been observed with other putative Titan materials (e.g., 1,3-butadiene, which also adopts a monoclinic structure).[48] The volume of the pyridine–acetylene unit cell expands by ∼2.5% (Figure 8), which is on par with previously characterized co-crystals.[57]

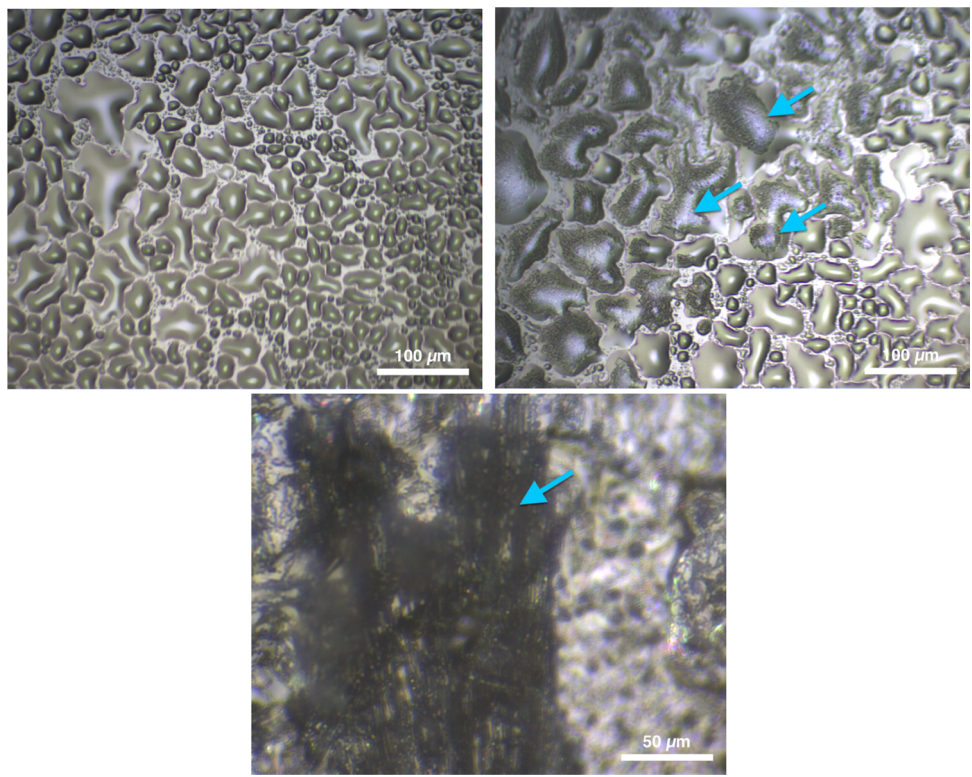

**Figure 5.** Top-down microscopic images depicting pyridine:acetylene (1:1) co-crystal formation. Top left: pyridine in the cryostage at 213 K (10× magnification). Top right: a mixture of pyridine and acetylene at 193 K. Examples of this mixed texture are indicated by arrows, although the texture is present throughout the image (10× magnification). The light-toned portion of the sample is pure pyridine, and the dark-toned portion of the sample is acetylene, which has crystallized within pyridine. Bottom: The co-crystal section of the sample at 163 K (50× magnification). Notice the relatively low albedo and "brainy" texture of the co-crystal (indicated by an arrow) compared to the surrounding sample.

## 5. CO-CRYSTAL STABILITY AFTER THE ETHANE WETTING EVENT

Titan raindrops are predicted to be primarily methane–nitrogen in composition, but as raindrops fall through the atmosphere, ethane content may increase after the droplet reaches compositional equilibrium.[49] Additionally, the altitude of observed cloud systems associated with Titan's lakes agrees with what may be expected for the winter subsidence of ethane.[50] Further, the Huygens probe found evidence of volatilized ethane at its landing site after touchdown.[51,52] As Titan rainstorms, and liquid ethane exposure in general (i.e., flowing liquid ethane), could alter the surface chemistry, stability, and duration of molecules in certain phases (i.e., co-crystals); thus, we have simulated a liquid ethane event in the XRD capillary to study how liquid ethane exposure could affect the pyridine:acetylene co-crystal. Liquid ethane was condensed inside the XRD capillary after co-crystal formation was verified (refer to methods in Section 2.2). Figure 7 shows the ethane wetting event pattern in blue (110 K) in comparison with the co-crystal pattern at the same temperature prior to exposure. This wetting event was carried out at 110 K to facilitate more rapid ethane evaporation on the timescale of these experiments. Note that the characteristic co-crystal peaks (e.g., at 11, 19.4, 20.1, 20.25°) are still observable immediately after ethane exposure. These features continue to persist after letting ethane interact with the sample for >20 h, suggesting stability over longer timescales than our experiment.

## 6. DISCUSSION

**6.1. Comparison with Previously Reported Co-crystals.** Considering several other Titan-relevant co-crystals have been formed and analyzed using similar techniques described herein (Table 5), it is important to compare their physical properties. The density of the pyridine:acetylene (1:1) co-crystal is 1.005 g/cm$^3$ (at 185 K),[33] which is most similar to the benzene:acetylene (1:1) co-crystal, reported at 1.009 g/cm$^3$ (Table 5).[53] We can infer that the similar densities between these co-crystals may be a result of the 1:1 stoichiometric ratio they have in common and the similar molecular weights of benzene and pyridine (78.11 and 79.1 g/mol, respectively). The acetylene:ammonia (1:1) co-crystal also shares the 1:1 stoichiometric ratio, albeit a higher density at 1.694 g/cm$^3$.[53] It is important to note that co-crystals such as pyridine bonded to two half-molecules of acetylene have longer C−H⋯N hydrogen bridge lengths (2.485 Å) compared to acetylene:ammonia (2.363 Å);[33] therefore, the acetylene:ammonia co-crystal exhibits denser packing and thus a higher density than the pyridine:acetylene co-crystal. Additionally, the ammonia molecule is smaller than pyridine, allowing for denser packing.

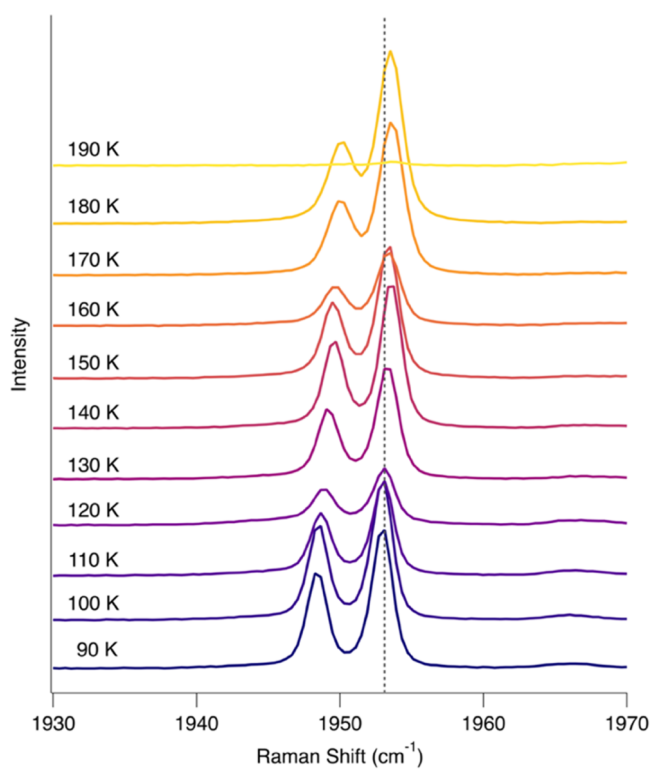

Figure 6. Thermal stability of the pyridine:acetylene co-crystal in the acetylene C≡C stretching region. Spectra are vertically offset and normalized for clarity. The co-crystal bands at 1948.3 and 1953.1 cm$^{-1}$ persist up to 180 K. These spectra show that the co-crystal is stable from 90 to 180 K.

When comparing Raman spectral features among co-crystals, we also observe similarities in band center positions. For example, in the acetylene C–C stretching region, the acetylene:ammonia co-crystal has features at 1944.4 cm$^{-1}$, which is comparable to the pyridine:acetylene co-crystal features at 1948.3 cm$^{-1}$. The pyridine:acetylene co-crystal also has peaks at 1953.1 and 1966 cm$^{-1}$ that are near the acetonitrile:acetylene, acetylene clathrate hydrate, and butane:acetylene peaks at 1957.1, 1966, and 1967.3 cm$^{-1}$, respectively. The commonality amongst these peaks is inferred to be a result of acetylene being a common co-former. Therefore, a Raman spectrometer that would characterize Titan's surface on a future in situ mission may need a spectral resolution better than ~4 cm$^{-1}$ to distinguish between spectral features and uniquely identify acetylene-bearing co-crystals, especially if these cryominerals are present as mixtures in surface materials. Further, Titan surface materials may prevent the identification of acetylene-bearing co-crystals with an in situ Raman spectrometer (spectral resolution better than ~4 cm$^{-1}$) if surface materials also have spectral features that overlap with those of acetylene-bearing co-crystals.

Table 4. Refined Lattice Constants and Unit Cell Volumes of the Pyridine:Acetylene Co-Crystal from 90–150 K, as Obtained from the Pawley Refinement of the Temperature-Series Data

| temperature (K) | $a$ (Å) | $b$ (Å) | $c$ (Å) | $\beta$ (deg) | volume (Å$^3$) |
|---|---|---|---|---|---|
| 90  | 5.8387 | 7.2757 | 16.0688 | 90.897 | 682.30 |
| 100 | 5.8457 | 7.2941 | 16.0845 | 90.853 | 685.75 |
| 110 | 5.8493 | 7.3133 | 16.1208 | 90.872 | 689.53 |
| 120 | 5.8553 | 7.3231 | 16.1522 | 90.910 | 692.51 |
| 130 | 5.8670 | 7.3227 | 16.1959 | 90.949 | 695.71 |
| 140 | 5.8702 | 7.3287 | 16.2156 | 90.922 | 697.52 |
| 150 | 5.8721 | 7.3346 | 16.2446 | 90.896 | 699.56 |

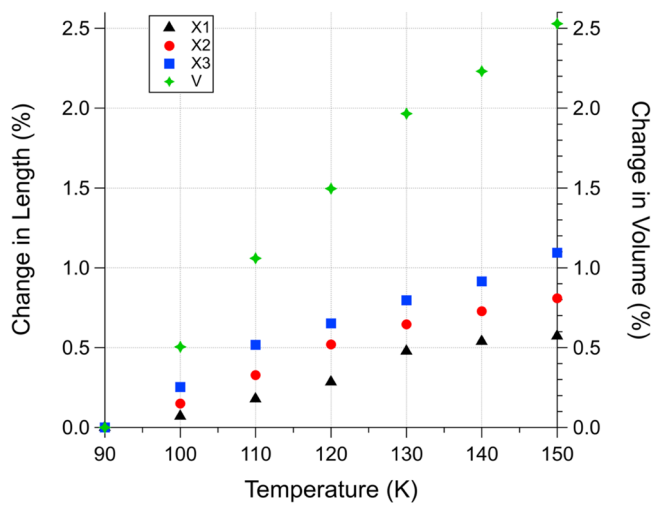

Figure 8. Percent change in the volume and length of the pyridine–acetylene unit cell along the principal axes (X1, X2, X3). Values are calculated from the refined lattice parameters in Table 4 using PASCal software.

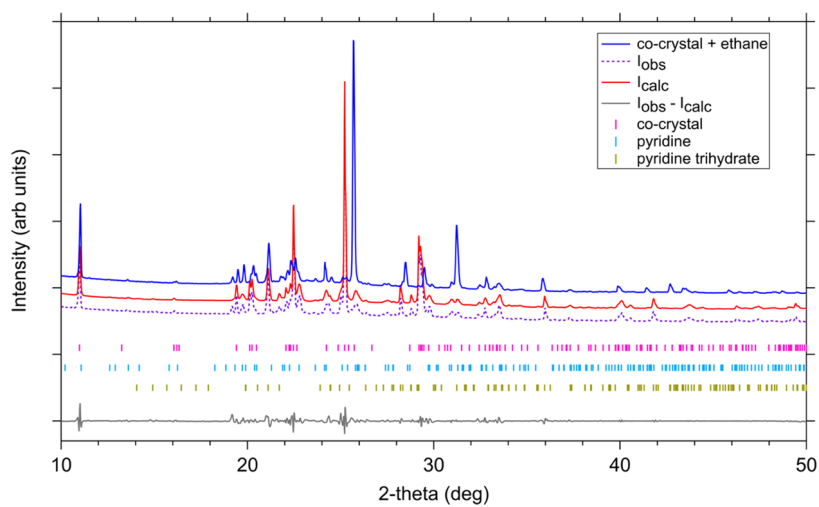

Figure 7. XRD pattern of the pyridine:acetylene co-crystal at 110 K (purple dash), the calculated Pawley refinement (red), and residual pattern (gray, offset for clarity). Tick marks below the patterns represent the Bragg peak positions of the co-crystal (magenta), pyridine (cyan), and pyridine trihydrate (gold). The co-crystal is most noticeable by the peak at 10.99°. The blue pattern shows the pyridine:acetylene co-crystal after an ethane wetting event at 110 K. Co-crystal peaks were still clearly detectable after letting ethane interact with the sample for >20 h, suggesting stability over longer timescales than our experiment. We note that the blue pattern is a different experiment from the red/purple one and therefore had different amounts of excess pyridine.

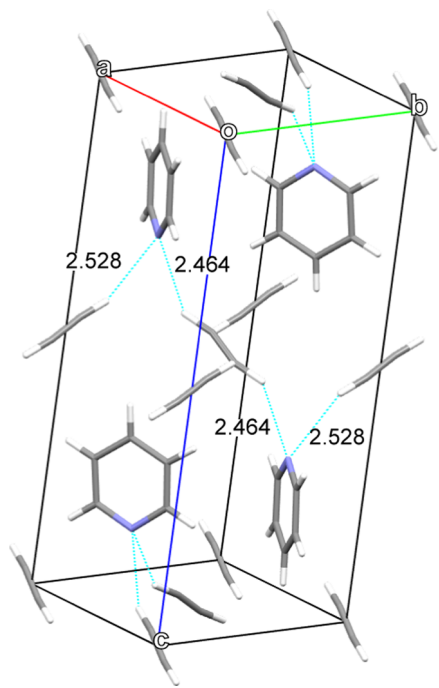

**Figure 9.** Primary intermolecular reactions that stabilize the pyridine:acetylene (1:1) co-crystal, represented by dashed cyan lines. Each pyridine N atom (blue) is bonded to the terminal H atoms (white) of two opposing acetylene molecules, with contact lengths labeled in Å, from the Cambridge Structural Database (CSD) Refcode WAFNIB, as determined by Kirchner et al.[33]

The anisotropic thermal expansion of the pyridine:acetylene co-crystal is common to multiple co-crystals. Anisotropic thermal expansion was observed with the acetonitrile:acetylene co-crystal from ~0.5 to 1% in all three axes; the $c$ axis was stabilized by strong N···H–C interactions from two acetylene molecules,[31] similar to what is observed here with the pyridine:acetylene co-crystal. Additionally, the benzene:ethane co-crystal expanded anisotropically along the $a$ and $b$ axes up to ~1%, which is explained by relatively weak C–H···$\pi$ interactions along the $a$ and $b$ axes compared to stronger, interlocking chains along the $c$ axis.[54] While the thermal expansion of the pyridine:acetylene co-crystal is similar to other acetylene co-crystal formers, we note that the acetylene:ammonia (1:1) co-crystal exhibits the most significant thermal expansion by far compared to any cryominerals reported to date.[54,58]

**6.2. Relevance to Geologic Processes on Titan.** During the ethane wetting experiment, all co-crystal peaks were still observable immediately after being exposed to liquid ethane (Figure 7). We note that the pyridine:acetylene (1:1) co-crystal was also stable after interacting with ethane for over 20 h at 110 K, suggesting potential stability on much longer timescales. In the context of Titan, co-crystals may provide a unique setting that allows certain compounds that are highly soluble in Titan liquids (e.g., acetylene solubility in ethane is 0.48 mole fraction[64]) to be preferentially "sequestered" as a molecular mineral. Further, it is likely that surface materials on Titan are complex mixtures comprised of additional organics, and while ternary co-crystals have been proposed,[61] these have yet to be confirmed experimentally. co-crystals may be tentatively detected on Titan's surface via NASA's *Dragonfly* mission, a rotorcraft lander that will provide in situ measurements of Titan's organic chemistry and habitability.[65][−67] Surface morphology (including microscale features) will be imaged by the camera system (DragonCam), the bulk elemental composition will be elucidated with the gamma-ray and neutron spectrometer (DraGNS) instrument, and more detailed molecular analysis including molecular ratios will be provided by the mass spectrometer (DraMS); combined, these instruments may be able to discern which cryominerals exist and are stable at the surface.

**Table 5. Previously Reported Titan-Relevant Co-Crystals, Temperature Stability, Formation Time, Detection Techniques, and Implications for Titan**[b]

| co-crystal | stability | density at ~100 K (g cm$^{-3}$)[55] | formation timescale | method(s) used | Titan implications |
|---|---|---|---|---|---|
| carbon dioxide:acetylene[56] | metastable | TBD | unknown; decomposes after a few minutes at 79 K | FTIR | likely to form in the troposphere. surface presence could indicate current deposition or CO$_2$ outgassing |
| benzene:ethane (3:1)[54,57,58] | <160 K | 1.067 | within minutes at 140 K | micro-Raman XRD | benzene-containing evaporites may not be pure species, but rather co-crystals. |
| acetylene:ammonia (1:1)[59,60] | <115 K | 1.694 | within minutes at 90 K | micro-Raman | may contribute to selective sequestration of ammonia |
| butane:acetylene[40] | <190 K | TBD | within minutes at 130 K | micro-Raman | butane-containing evaporites may not be pure species, but rather co-crystals |
| benzene:acetylene:hydrogen cyanide (2:1:1)[61] | TBD | 1.913[a] | TBD | DFT | implies the formation of complex organics in the atmosphere |
| acetonitrile:acetylene (1:2)[31] | <120 K <170 K | 1.260 | within minutes at 90 K | micro-Raman XRD | possible component of labyrinth terrains |
| benzene:acetonitrile (3:1)[62] | <245 K | 1.096 | TBD | XRD | potential phase change upon liquid C$_2$H$_6$ exposure; may indicate past C$_2$H$_6$ presence |
| benzene:acetylene (1:1)[63] | <135 K | 1.009 | within minutes at 135 K | FTIR | complex co-crystallization could occur in the atmosphere |

[a]Calculated. [b]DFT: density functional theory.

Because of the relatively large amount of acetylene predicted on Titan's surface compared to the estimated abundance of pyridine, it is possible that the majority of pyridine on Titan may be preferentially concentrated in the form of the co-crystal. The co-crystal is most easily formed from a liquid phase (at least under experimental timescales), which suggests that warmer environments or liquid interactions may be conducive for this co-crystal to form in situ on Titan. We note that temperatures in Titan's stratosphere reach and exceed 150 K,[36] so it is possible that acetylene could come into contact with liquid pyridine as an aerosol in the atmosphere. If present at Titan's surface, both pyridine and acetylene would exist in their solid phases. Initial experiments testing for solid−solid co-crystal formation between pyridine and acetylene were unsuccessful (previously reported co-crystals have formed via solid−solid interactions, e.g., benzene:acetylene[63]). The experimental condition for the pyridine:acetylene (1:1) co-crystal to form was most readily achieved with liquid pyridine (liquid phase from 231.6 to 388 K); the temperature range at which acetylene is in the liquid phase is relatively narrow (approx. 193 to 189 K). Thus, assuming that the pyridine:acetylene co-crystal is identified on Titan's surface, one could infer that pyridine may have existed in the liquid phase on the surface in the past. Although Titan's average surface temperature is ∼90 K, localized energetic events (i.e., cryovolcanism, impact cratering) could allow surface temperatures in excess of 200 K.[68] Further, thermal modeling by Neish et al. suggests that liquid water or water−ammonia environments associated with cryovolcanism could be sustained for timescales on the order of $10^2$−$10^5$ years,[68] providing a potentially favorable environment for prebiotic molecules or co-crystals (i.e., the pyridine:acetylene co-crystal) to form and interact. Additionally, HCN (a significant prebiotic molecule that has been observed on Titan) may be available to dissolve in the liquid "cryomagma" either to yield more complex biomolecules (e.g., amino acids) or to combine with polymerized acetylene to yield pyridine production.

Another possibility is that the pyridine:acetylene co-crystal could form in Titan's warmer interior, which may reach temperatures in excess of 255 K. In that respect, this co-crystal may serve as our first example of a metamorphic cryomineral (i.e., may have been processed at higher temperatures/pressure below the surface). If it is discovered on the surface, that may be indicative of an area on Titan that has exposed material transported (or excavated) from the moon's interior. Possible mechanisms of transport from deeper zones include impact cratering[69] and laccolithic emplacement at depth (possibly with a second laccolithic emplacement that could lift the previous uplift even higher) that would lift (successively, perhaps) deeper areas of crust towards Titan's surface.[70] The Titan labyrinth terrains suggest that at least 500 m of throw is possible[71] and proposed mountain belts could result from methane-lubricated thrust faults that would result in uplift.[72]

Considering that the entire sample did not become co-crystalline—only certain localized areas—we may also expect to observe co-crystal features within patches of pure acetylene and pyridine on Titan's surface. This "patchiness" may have occurred in the experiments because of relatively short reaction times compared to Titan geologic timescales. Longer timescales might allow for the co-crystal to form at lower temperatures under Titan surface conditions (89−94 K) or more homogeneously in surface materials, even though that cannot be reproduced on experimental timescales. At lower temperatures, mixtures of pure pyridine and acetylene were observed in our experiments. Mixtures like this are common in our experiments, as there are a variety of factors that may prevent the "ideal" stoichiometry of pyridine and acetylene from being met across the entire sample area. Some of these include temperature variation across the slide, diffusion, or rate of acetylene deposition with respect to pyridine freezing. These are just a few of the many examples of why a pure co-crystalline sample is not expected to form. A kinetics study would be needed to determine how quickly the co-crystal forms as a function of temperature, but that is beyond the scope of this paper. Further, the physical processing of a heterogenous mixture of acetylene and pyridine could either produce the co-crystal or redistribute the pure compounds where they may have the chance to react further with other compounds. For example, the pyridine:acetylene co-crystal (or pure components) may be transported to or formed in Titan's subsurface where warmer temperatures may allow contact with liquid water (or ammonia−water liquids[35,73]) and potential access to putative life. Co-crystals allow for the concentration and increased stabilization of acetylene, even after exposure to liquid ethane. Thus, if acetylene-rich deposits exist on Titan's surface and interact with N-heterocycles like pyridine, these interactions could concentrate ingredients that may be needed to support putative life.

We note that the pyridine:acetylene (1:1) co-crystal exists and is stable under Titan-relevant conditions in our lab experiments, where these ideal conditions are created; however, longer timescale geologic processes that actually exist on Titan are unable to be tested for in a laboratory environment. Additionally, there are still many unknowns regarding the exact composition of Titan's surface (many of these will be addressed by future missions, such as *Dragonfly*). We provide these laboratory measurements for the ideal case where such conditions and interactions may be observed on Titan in the future.

## 7. CONCLUSIONS

We have shown that the pyridine:acetylene (1:1) co-crystal forms readily at 150 K and is stable from 90−180 K. The co-crystal is durable in the case of an ethane "wetting" event, simulating fluvial/pluvial interactions that may occur on Titan. Similar to previously reported co-crystals and putative Titan solids, the pyridine:acetylene (1:1) co-crystal exhibits anisotropic thermal expansion over the temperature range studied. Additionally, the pyridine:acetylene (1:1) co-crystal shares peak positions with other acetylene-formed co-crystals, which underscores the need for acquiring in situ, high-resolution compositional data from Titan's surface. Although only upper limits of pyridine in Titan's atmosphere have been predicted, the high abundance of acetylene on Titan may allow any pyridine present to preferentially sequester into the co-crystal form. Further, the presence of the pyridine:acetylene (1:1) co-crystal on Titan (if detected) may infer warmer surface temperatures in the past or be associated with geologic processes such as cryovolcanism, impact cratering, or subsurface processing/transport. In general, co-crystals with astrobiologically relevant molecules (i.e., acetylene and pyridine) allow for the concentration of prebiotic ingredients and energy sources that may facilitate putative life. Future studies will characterize more complex co-crystals such as ternary systems and those with other nitrile species, which will further elucidate this growing field of cryomineralogy on Titan.

## ASSOCIATED CONTENT

**Data Availability Statement**

Data on the pyridine:acetylene co-crystal, including Raman spectra and XRD patterns, can be found at https://doi.org/10.48577/jpl.1UHZFY.

**Supporting Information**

The Supporting Information is available free of charge at https://pubs.acs.org/doi/10.1021/acsearthspacechem.2c00377.

> Schematic diagrams of the Raman and XRD experimental setups (Figures S1 and S2); experimental Raman shifts of the co-crystal lattice vibrational modes (Table S1 and Figure S3), and Raman spectra of the pyridine trihydrate (Figure S4) (PDF)

## AUTHOR INFORMATION

**Corresponding Author**

Ellen C. Czaplinski − *NASA Jet Propulsion Laboratory, California Institute of Technology, Pasadena, California 91109, United States;* orcid.org/0000-0002-2046-1416; Email: ellen.c.czaplinski@jpl.nasa.gov

**Authors**

Tuan H. Vu − *NASA Jet Propulsion Laboratory, California Institute of Technology, Pasadena, California 91109, United States;* orcid.org/0000-0001-6839-9765

Morgan L. Cable − *NASA Jet Propulsion Laboratory, California Institute of Technology, Pasadena, California 91109, United States;* orcid.org/0000-0002-3680-302X

Mathieu Choukroun − *NASA Jet Propulsion Laboratory, California Institute of Technology, Pasadena, California 91109, United States;* orcid.org/0000-0001-7447-9139

Michael J. Malaska − *NASA Jet Propulsion Laboratory, California Institute of Technology, Pasadena, California 91109, United States*

Robert Hodyss − *NASA Jet Propulsion Laboratory, California Institute of Technology, Pasadena, California 91109, United States*

Complete contact information is available at:
https://pubs.acs.org/10.1021/acsearthspacechem.2c00377

**Notes**

The authors declare no competing financial interest.

## ACKNOWLEDGMENTS

This research was supported by appointment to the NASA Postdoctoral Program at the Jet Propulsion Laboratory administered by Oak Ridge Associated Universities under contract with NASA. This work was conducted at the Jet Propulsion Laboratory, California Institute of Technology, under a contract with the National Aeronautics and Space Administration (80NM0018D0004). Reference herein to any specific commercial product, process, or service by trade name, trademark, manufacturer, or otherwise does not constitute or imply its endorsement by the United States Government or the Jet Propulsion Laboratory, California Institute of Technology. Government sponsorship is acknowledged. 2023. All rights reserved.

## REFERENCES

(1) Krasnopolsky, V. A. A Photochemical Model of Titan's Atmosphere and Ionosphere. *Icarus* **2009**, *201*, 226−256.
(2) Willacy, K.; Allen, M.; Yung, Y. A New Astrobiological Model of the Atmosphere of Titan. *Astrophys. J.* **2016**, *829*, No. 79.
(3) Hörst, S. M. Titan's Atmosphere and Climate. *J. Geophys. Res.: Planets* **2017**, *122*, 432−482.
(4) Lavvas, P. P.; Coustenis, A.; Vardavas, I. M. Coupling Photochemistry with Haze Formation in Titan's Atmosphere, Part I: Model Description. *Planet. Space Sci.* **2008**, *56*, 27−66.
(5) Loison, J. C.; Dobrijevic, M.; Hickson, K. M. The Photochemical Production of Aromatics in the Atmosphere of Titan. *Icarus* **2019**, *329*, 55−71.
(6) Krasnopolsky, V. A. Chemical Composition of Titan's Atmosphere and Ionosphere: Observations and the Photochemical Model. *Icarus* **2014**, *236*, 83−91.
(7) Vuitton, V.; Yelle, R. V.; McEwan, M. J. Ion Chemistry and N-Containing Molecules in Titan's Upper Atmosphere. *Icarus* **2007**, *191*, 722−742.
(8) Lara, L. M.; Lellouch, E.; López-Moreno, J. J.; Rodrigo, R. Vertical Distribution of Titan's Atmospheric Neutral Constituents. *J. Geophys. Res.: Planets* **1996**, *101*, 23261−23283.
(9) Coustenis, A.; Achterberg, R. K.; Conrath, B. J.; Jennings, D. E.; Marten, A.; Gautier, D.; Nixon, C. A.; Flasar, F. M.; Teanby, N. A.; Bézard, B.; Samuelson, R. E.; Carlson, R. C.; Lellouch, E.; Bjoraker, G. L.; Romani, P. N.; Taylor, F. W.; Irwin, P. G. J.; Fouchet, T.; Hubert, A.; Orton, G. S.; Kunde, V. G.; Vinatier, S.; Mondellini, J.; Abbas, M. M.; Courtin, R. The Composition of Titan's Stratosphere from Cassini/CIRS Mid-Infrared Spectra. *Icarus* **2007**, *189*, 35−62.
(10) Singh, S.; Mccord, T. B.; Combe, J.; Rodriguez, S.; Cornet, T.; Le Mouélic, S.; Clark, R. N.; Maltagliati, L.; Chevrier, V. F. Acetylene on Titan's Surface. *Astrophys. J.* **2016**, *828*, No. 55.
(11) Waite, J. H., Jr.; Niemann, H.; Yelle, R. V.; Kasprzak, W. T.; Cravens, T. E.; Luhmann, J. G.; McNutt, R. L.; Ip, W.-H.; Gell, D.; de La Haye, V.; Müller-Wordag, I.; Magee, B.; Borggren, N.; Ledvina, S.; Fletcher, G.; Walter, E.; Miller, R.; Scherer, S.; Thorpe, R.; Xu, J.; Block, B.; Arnett, K. Ion Neutral Mass Spectrometer Results from the First Flyby of Titan. *Science* **2005**, *308*, 982−986.
(12) Bottger, G. L.; Eggers, D. F. Infrared Spectra of Crystalline C2H2, C2HD, and C2D2. *J. Chem. Phys.* **1964**, *40*, 2010−2017.
(13) Stoks, P. G.; Schwartz, A. W. Nitrogen-Heterocyclic Compounds in Meteorites: Significance and Mechanisms of Formation. *Geochim. Cosmochim. Acta* **1981**, *45*, 563−569.
(14) Stoks, P. G.; Schwartz, A. W. Basic Nitrogen-Heterocyclic Compounds in the Murchison Meteorite. *Geochim. Cosmochim. Acta* **1982**, *46*, 309−315.
(15) Cronin, J. R.; Change, S.Organic Matter in Meteorites: Molecular and Isotopic Analyses of the Murchison Meteorite. In *The Chemistry of Life's Origins*; Springer, 1993; pp 209−258.
(16) Botta, O.; Bada, J. L. Extraterrestrial Organic Compounds in Meteorites. *Surv. Geophys.* **2002**, *23*, 411−467.
(17) Sephton, M. A. Organic Compounds in Carbonaceous Meteorites. *Nat. Prod. Rep.* **2002**, *19*, 292−311.
(18) Charnley, S. B.; Kuan, Y. J.; Huang, H. C.; Botta, O.; Butner, H. M.; Cox, N.; Despois, D.; Ehrenfreund, P.; Kisiel, Z.; Lee, Y. Y.; Markwick, A. J.; Peeters, Z.; Rodgers, S. D. Astronomical Searches for Nitrogen Heterocycles. *Adv. Space Res.* **2005**, *36*, 137−145.
(19) Hudson, R. L.; Yarnall, Y. Y. Infrared Spectra and Optical Constants of Astronomical Ices: IV. Benzene and Pyridine. *Icarus* **2022**, *377*, No. 114899.

(20) Paubert, G.; Marten, A.; Rosolen, C.; Gautier, D.; Courtin, R. First Radiodetection of HCN on Titan. *Bull. Am. Astron. Soc.* **1987**, *19*, No. 633.

(21) Hidayat, T.; Marten, A.; Bézard, B.; Gautier, D.; Owen, T.; Matthews, H. E.; Paubert, G. Millimeter and Submillimeter Heterodyne Observations of Titan: Retrieval of the Vertical Profile of HCN and the 12C/13C Ratio. *Icarus* **1997**, *126*, 170−182.

(22) Marten, A.; Hidayat, T.; Biraud, Y.; Moreno, R. New Millimeter Heterodyne Observations of Titan: Vertical Distributions of Nitriles HCN, HC3N, CH3CN, and the Isotopic Ratio 15N/14N in Its Atmosphere. *Icarus* **2002**, *158*, 532−544.

(23) Courtin, R.; Swinyard, B. M.; Moreno, R.; Fulton, T.; Lellouch, E.; Rengel, M.; Hartogh, P. First Results of Herschel-SPIRE Observations of Titan. *Astron. Astrophys.* **2011**, *536*, No. L2.

(24) Rengel, M.; Shulyak, D.; Hartogh, P.; Sagawa, H.; Moreno, R.; Jarchow, C.; Breitschwerdt, D. Ground-Based HCN Submillimetre Measurements in Titan's Atmosphere: An Intercomparison with Herschel Observations. *Astron. Astrophys.* **2022**, *658*, No. A88.

(25) Ricca, A.; Bauschlicher, C. W.; Bakes, E. L. O. A Computational Study of the Mechanisms for the Incorporation of a Nitrogen Atom into Polycyclic Aromatic Hydrocarbons in the Titan Haze. *Icarus* **2001**, *154*, 516−521.

(26) Soorkia, S.; Taatjes, C. A.; Osborn, D. L.; Selby, T. M.; Trevitt, A. J.; Wilson, K. R.; Leone, S. R. Direct Detection of Pyridine Formation by the Reaction of CH (CD) with Pyrrole: A Ring Expansion Reaction. *Phys. Chem. Chem. Phys.* **2010**, *12*, 8750−8758.

(27) Cerceau, F.; Raulin, F.; Courtin, R.; Gautier, D. Infrared Spectra of Gaseous Mononitriles: Application to the Atmosphere of Titan. *Icarus* **1985**, *62*, 207−220.

(28) Cui, J.; Yelle, R. V.; Vuitton, V.; Waite, J. H.; Kasprzak, W. T.; Gell, D. A.; Niemann, H. B.; Müller-Wodarg, I. C. F.; Borggren, N.; Fletcher, G. G.; Patrick, E. L.; Raaen, E.; Magee, B. A. Analysis of Titan's Neutral Upper Atmosphere from Cassini Ion Neutral Mass Spectrometer Measurements. *Icarus* **2009**, *200*, 581−615.

(29) Nixon, C. A.; Thelen, A. E.; Cordiner, M. A.; Kisiel, Z.; Charnley, S. B.; Molter, E. M.; Serigano, J.; Irwin, P. G. J.; Teanby, N. A.; Kuan, Y.-J. Detection of Cyclopropenylidene on Titan with ALMA. *Astron. J.* **2020**, *160*, No. 205.

(30) Bond, A. D. What Is a Co-Crystal? *CrystEngComm* **2007**, *9*, 833−834.

(31) Cable, M. L.; Vu, T. H.; Malaska, M. J.; Maynard-Casely, H. E.; Choukroun, M.; Hodyss, R. Properties and Behavior of the Acetonitrile-Acetylene Co-Crystal under Titan Surface Conditions. *ACS Earth Space Chem.* **2020**, *4*, 1375−1385.

(32) Stang, P. J.; Diederich, F. *Modern Acetylene Chemistry*; VCH Weinheim: New York, 1995.

(33) Kirchner, M. T.; Bläser, D.; Boese, R. Co-Crystals with Acetylene: Small Is Not Simple! *Chem. - Eur. J.* **2010**, *16*, 2131−2146.

(34) Sohl, F.; Hussmann, H.; Schwentker, B.; Spohn, T.; Lorenz, R. D. Interior Structure Models and Tidal Love Numbers of Titan. *J. Geophys. Res. Planets* **2003**, *108* (E12), 5130.

(35) Sohl, F.; Solomonidou, A.; Wagner, F. W.; Coustenis, A.; Hussmann, H.; Schulze-Makuch, D. Structural and Tidal Models of Titan and Inferences on Cryovolcanism. *J. Geophys. Res. Planets* **2014**, *119*, 1013−1036.

(36) Flasar, F. M.; Achterberg, R. K.; Conrath, B. J.; Gierasch, P. J.; Kunde, V. G.; Nixon, C. A.; Bjoraker, G. L.; Jennings, D. E.; Romani, P. N.; Simon-Miller, A. A.; Bézard, B.; Coustenis, A.; Irwin, P. G. J.; Teanby, N. A.; Brasunas, J.; Pearl, J. C.; Segura, M. E.; Carlson, R. C.; Mamoutkine, A.; Schinder, P. J.; Barucci, A.; Courtin, R.; Fouchet, T.; Gautier, D.; Lellouch, E.; Marten, A.; Prangé, R.; Vinatier, S.; Strobel, D. F.; Calcutt, S. B.; Read, P. L.; Taylor, F. W.; Bowles, N.; Samuelson, R. E.; Orton, G. S.; Spilker, L. J.; Owen, T. C.; Spencer, J. R.; Showalter, M. R.; Ferrari, C.; Abbas, M. M.; Raulin, F.; Edgington, S.; Ade, P.; Wishnow, E. H. Titan's Atmospheric Temperatures, Winds, and Composition. *Science* **2005**, *308*, 975−979.

(37) Mcintosh, D. The Physical Properties of Liquid and Solid Acetylene. *J. Phys. Chem. A* **1907**, *11*, 306−317.

(38) Vuitton, V.; Yelle, R. V.; Klippenstein, S. J.; Hörst, S. M.; Lavvas, P. Simulating the Density of Organic Species in the Atmosphere of Titan with a Coupled Ion-Neutral Photochemical Model. *Icarus* **2019**, *324*, 120−197.

(39) Hodyss, R.; Vu, T. H.; Choukroun, M.; Cable, M. L. A Simple Gas Introduction System for Cryogenic Powder X-Ray Diffraction. *J. Appl. Crystallogr.* **2021**, *54*, 1268−1270.

(40) Cable, M. L.; Vu, T. H.; Malaska, M. J.; Maynard-Casely, H. E.; Choukroun, M.; Hodyss, R. A Co-Crystal between Acetylene and Butane: A Potentially Ubiquitous Molecular Mineral on Titan. *ACS Earth Space Chem.* **2019**, *3*, 2808−2815.

(41) Vu, T. H.; Hodyss, R.; Cable, M. L.; Choukroun, M. Raman Signatures and Thermal Expansivity of Acetylene Clathrate Hydrate. *J. Phys. Chem. A* **2019**, *123*, 7051−7056.

(42) Anderson, A.; Andrews, B.; Torrie, B. H. Raman and Far Infrared Spectra of Crystalline Acetylene, C2H2 and C2D2. *J. Raman Spectrosc.* **1985**, *16*, 202−207.

(43) Hey, A. M.; Venter, M. W. Effect of Pressure on the Raman Spectra of Solids. 2. Pyridine. 1985, https://pubs.acs.org/sharingguidelines.

(44) Castellucci, E.; Sbrana, G.; Verderame, F. D. Infrared Spectra of Crystalline and Matrix Isolated Pyridine and Pyridine-D5. *J. Chem. Phys.* **1969**, *51*, 3762−3770.

(45) Sugawara, T.; Kanda, E. The Crystal Structure of Acetylene. *Sci. Rep. Res. Inst., Tohoku Univ. Ser. A* **1952**, *4*, 607−614.

(46) Mootz, D.; Wussow, H.-G. Crystal Structures of Pyridine and Pyridine Trihydrate. *J. Chem. Phys.* **1981**, *75*, 1517−1522.

(47) Cliffe, M. J.; Goodwin, A. L. PASCal: A Principal Axis Strain Calculator for Thermal Expansion and Compressibility Determination. *J. Appl. Crystallogr.* **2012**, *45*, 1321−1329.

(48) Vu, T. H.; Maynard-Casely, H. E.; Cable, M. L.; Choukroun, M.; Malaska, M. J.; Hodyss, R. 1,3-Butadiene on Titan: Crystal Structure, Thermal Expansivity, and Raman Signatures. *ACS Earth Space Chem.* **2022**, *6*, 2274−2281.

(49) Dalba, P. A.; Buratti, B. J.; Brown, R. H.; Barnes, J. W.; Baines, K. H.; Sotin, C.; Clark, R. N.; Lawrence, K. J.; Nicholson, P. D. Cassini VIMS Observations Show Ethane Is Present in Titan's Rainfall. *Astrophys. J.* **2012**, *761*, No. L24.

(50) Griffith, C. A.; Penteado, P.; Rannou, P.; Brown, R.; Boudon, V.; Baines, K. H.; Clark, R.; Drossart, P.; Buratti, B.; Nicholson, P.; McKay, C. P.; Coustenis, A.; Negrao, A.; Jaumann, R. Evidence for a Polar Ethane Cloud on Titan. *Science* **2006**, *313*, 1620−1623.

(51) Niemann, H. B.; Atreya, S. K.; Bauer, S. J.; Carignan, G. R.; Demick, J. E.; Frost, R. L.; Gautier, D.; Haberman, J. A.; Harpold, D. N.; Hunten, D. M.; Israel, G.; Lunine, J. I.; Kasprzak, W. T.; Owen, T. C.; Paulkovich, M.; Raulin, F.; Raaen, E.; Way, S. H. The Abundances of Constituents of Titan's Atmosphere from the GCMS Instrument on the Huygens Probe. *Nature* **2005**, *438*, 779−784.

(52) Lorenz, R. D.; Niemann, H. B.; Harpold, D. N.; Way, S. H.; Zarnecki, J. C. Titan's Damp Ground: Constraints on Titan Surface Thermal Properties from the Temperature Evolution of the Huygens GCMS Inlet. *Meteorit. Planet. Sci.* **2006**, *41*, 1705−1714.

(53) Boese, R.; Blaser, D.; Jansen, G. Synthesis and Theoretical Characterization of an Acetylene-Ammonia co-crystal. *J. Am. Chem. Soc.* **2009**, *131*, 2104−2106.

(54) Maynard-Casely, H. E.; Hodyss, R.; Cable, M. L.; Vu, T. H.; Rahm, M. A Co-Crystal between Benzene and Ethane: A Potential Evaporite Material for Saturn's Moon Titan. *IUCrJ* **2016**, *3*, 192−199.

(55) Cable, M. L.; Runčevski, T.; Maynard-Casely, H. E.; Vu, T. H.; Hodyss, R. Titan in a Test Tube: Organic Co-Crystals and Implications for Titan Mineralogy. *Acc. Chem. Res.* **2021**, *54*, 3050−3059.

(56) Gough, T. E.; Rowat, T. E. Measurements of the Infrared Spectra and Vapor Pressure of the System Carbon Dioxide-Acetylene at Cryogenic Temperatures. *J. Chem. Phys.* **1998**, *109*, 6809−6813.

(57) Vu, T. H.; Cable, M. L.; Choukroun, M.; Hodyss, R.; Beauchamp, P. Formation of a New Benzene-Ethane Co-Crystalline Structure under Cryogenic Conditions. *J. Phys. Chem. A* **2014**, *118*, 4087−4094.

(58) Cable, M. L.; Vu, T. H.; Hodyss, R.; Choukroun, M.; Malaska, M. J.; Beauchamp, P. Experimental Determination of the Kinetics of Formation of the Benzene-Ethane Co-Crystal and Implications for Titan. *Geophys. Res. Lett.* **2014**, *41*, 5396−5401.

(59) Cable, M. L.; Vu, T. H.; Maynard-Casely, H. E.; Choukroun, M.; Hodyss, R. The Acetylene-Ammonia Co-Crystal on Titan. *ACS Earth Space Chem.* **2018**, *2*, 366−375.

(60) Vu, T. H.; Casely, H. E. M.; Cable, M. L.; Hodyss, R.; Choukroun, M.; Malaska, M. J. Anisotropic Thermal Expansion of the Acetylene-Ammonia Co-Crystal under Titan's Conditions. *J. Appl. Crystallogr.* **2020**, *53*, 1524−1530.

(61) Ennis, C.; Cable, M. L.; Hodyss, R.; Maynard-Casely, H. E. Mixed Hydrocarbon and Cyanide Ice Compositions for Titan's Atmospheric Aerosols: A Ternary-Phase Co-Crystal Predicted by Density Functional Theory. *ACS Earth Space Chem.* **2020**, *4*, 1195−1200.

(62) Mcconville, C. A.; Tao, Y.; Evans, H. A.; Trump, B. A.; Lefton, J. B.; Xu, W.; Yakovenko, A. A.; Kraka, E.; Brown, C. M.; Runčevski, T. Peritectic Phase Transition of Benzene and Acetonitrile into a co-crystal Relevant to Titan, Saturn's Moon. *Chem. Commun.* **2020**, *56*, 13520−13523.

(63) Czaplinski, E.; Yu, X.; Dzurilla, K.; Chevrier, V. Experimental Investigation of the Acetylene−Benzene Cocrystal on Titan. *Planet. Sci. J.* **2020**, *1*, No. 76.

(64) Singh, S.; Combe, J. P.; Cordier, D.; Wagner, A.; Chevrier, V. F.; McMahon, Z. Experimental Determination of Acetylene and Ethylene Solubility in Liquid Methane and Ethane: Implications to Titan's Surface. *Geochim. Cosmochim. Acta* **2017**, *208*, 86−101.

(65) Lorenz, R. D.; Turtle, E. P.; Barnes, J. W.; Trainer, M. G.; Adams, D. S.; Hibbard, K. E.; Sheldon, C. Z.; Zacny, K.; Peplowski, P. N.; Lawrence, D. J.; Ravine, M. A.; Mcgee, T. G.; Sotzen, K. S.; Mackenzie, S. M.; Langelaan, J. W.; Schmitz, S.; Wolfarth, L. S.; Bedini, P. D. *Dragonfly*: A Rotorcraft Lander Concept for Scientific Exploration at Titan. *APL Tech. Dig.* **2018**, *34*, 374−387.

(66) Turtle, E. P.; Barnes, J. W.; Trainer, M. G.; Lorenz, R. D.; Hibbard, K. E.; Adams, D. S.; Bedini, P.; Brinckerhoff, W. B.; Ernst, C.; Freissinet, C.; Hand, K.; Hayes, A. G.; Johnson, J. R.; Karkoschka, E.; Langelaan, J. W.; Gall, A. le.; Lora, J. M.; Mackenzie, S. M.; Mckay, C. P.; Neish, C. D.; Newman, C. E.; Palacios, J.; Parsons, A. M.; Peplowski, P. N.; Radebaugh, J.; Rafkin, S. C. R.; Ravine, M. A.; Schmitz, S.; Soderblom, J. M.; Sotzen, S.; Stickle, A. M.; Stofan, E. R.; Tokano, T.; Wilson, C.; Yingst, R. A.; Zacny, K.; Hopkins, J.; Physics, A.; Spatiales, O.; Angeles, L.; Field, M.; Space, M.; Systems, S.; Diego, S.; Robotics, H. *Dragonfly: In Situ Exploration of Titan's Organic Chemistry and Habitability*, 49th Annual Lunar and Planetary Science Conference, 2018; p 1641.

(67) Barnes, J. W.; Turtle, E. P.; Trainer, M. G.; Lorenz, R. D.; MacKenzie, S. M.; Brinckerhoff, W. B.; Cable, M. L.; Ernst, C. M.; Freissinet, C.; Hand, K. P.; Hayes, A. G.; Hörst, S. M.; Johnson, J. R.; Karkoschka, E.; Lawrence, D. J.; Le Gall, A.; Lora, J. M.; McKay, C. P.; Miller, R. S.; Murchie, S. L.; Neish, C. D.; Newman, C. E.; Núñez, J.; Panning, M. P.; Parsons, A. M.; Peplowski, P. N.; Quic, L. C.; Radebaugh, J.; Rafkin, S. C. R.; Shiraishi, H.; Soderblom, J. M.; Sotzen, K. S.; Stickle, A. M.; Stofan, E. R.; Szopa, C.; Tokano, T.; Wagner, T.; Wilson, C.; Yingst, R. A.; Zacny, K.; Stähler, S. C. Science Goals and Objectives for the *Dragonfly* Titan Rotorcraft Relocatable Lander. *Planet. Sci. J.* **2021**, *2*, No. 130.

(68) Neish, C. D.; Lorenz, R. D.; O'Brien, D. P. The Potential for Prebiotic Chemistry in the Possible Cryovolcanic Dome Ganesa Macula on Titan. *Int. J. Astrobiol.* **2006**, *5*, 57−65.

(69) Crósta, A. P.; Silber, E. A.; Lopes, R. M. C.; Johnson, B. C.; Bjonnes, E.; Malaska, M. J.; Vance, S. D.; Sotin, C.; Solomonidou, A.; Soderblom, J. M. Modeling the Formation of Menrva Impact Crater on Titan: Implications for Habitability. *Icarus* **2021**, *370*, No. 114679.

(70) Schurmeier, L. R.; Dombard, A. J.; Malaska, M. J.; Fagents, S. A.; Radebaugh, J.; Lalich, D. E. An Intrusive Cryomagmatic Origin for Northern Radial Labyrinth Terrains on Titan and Implications for the Presence of Crustal Clathrates. In Review at Icarus.

(71) Malaska, M. J.; Radebaugh, J.; Lopes, R. M. C.; Mitchell, K. L.; Verlander, T.; Schoenfeld, A. M.; Florence, M. M.; le Gall, A.; Solomonidou, A.; Hayes, A. G.; Birch, S. P. D.; Janssen, M. A.; Schurmeier, L.; Cornet, T.; Ahrens, C.; Farr, T. G.; Team, C. R. Labyrinth Terrain on Titan. *Icarus* **2020**, *344*, No. 113764.

(72) Liu, Z. Y.-C.; Radebaugh, J.; Harris, R. A.; Christiansen, E. H.; Neish, C. D.; Kirk, R. L.; Lorenz, R. D. The Tectonics of Titan: Global Structural Mapping from Cassini RADAR. *Icarus* **2016**, *270*, 14−29.

(73) Choukroun, M.; Grasset, O. Thermodynamic Data and Modeling of the Water and Ammonia-Water Phase Diagrams up to 2.2 GPa for Planetary Geophysics. *J. Chem. Phys.* **2010**, *133*, No. 144502.